# TRANSPORT PROPERTIES OF A GaAs/InGaAs/GaAs QUANTUM WELL : TEMPERATURE, MAGNETIC FIELD AND MANY-BODY EFFECTS

TRUONG VAN TUAN [a,b], NGUYEN QUOC KHANH [a,†], VO VAN TAI [a], AND DANG KHANH LINH [c]

[a] *University of Science - VNUHCM, 227-Nguyen Van Cu Street, 5th District, Ho Chi Minh City, Viet Nam*
[b] *University of Tran Dai Nghia, 189-Nguyen Oanh Street, Go Vap District, Ho Chi Minh City, Viet Nam*
[c] *Ho Chi Minh City University of Education, 280 An Duong Vuong Street, 5th District, Ho Chi Minh City, Vietnam*

[†] *E-mail:* nqkhanh@hcmus.edu.vn

**Abstract.** *We investigate the zero and finite temperature transport properties of a quasi-two-dimensional electron gas in a GaAs/InGaAs/GaAs quantum well under a magnetic field, taking into account many-body effects via a local-field correction. We consider the surface roughness, roughness-induced piezoelectric, remote charged impurity and homogenous background charged impurity scattering. The effects of the quantum well width, carrier density, temperature and local-field correction on resistance ratio are investigated. We also consider the dependence of the total mobility on the multiple scattering effect.*

## I. INTRODUCTION

GaAs/InGaAs/GaAs lattice-mismatched quantum well (QW) structure with a finite barrier height has been studied by several authors [1-4]. To asses the quality of new materials or electronic devices one needs to study their physical properties, among which transport ones such as mobility and resistivity turn out to be very important. In order to determine the main scattering mechanisms, which limit the mobility, one often compares theoretical results with those of obtained by experiments. Recently, Quang and his co-workers [5] have proposed a new scattering mechanism so called roughness-induced piezoelectric scattering and have calculated the zero-temperature mobility limited by this and surface roughness scattering for a GaAs/InGaAs/GaAs QW under zero magnetic field. By incorporating this scattering, they have explained successfully the low-temperature mobility measured for InGaAs-based QW's. Because electronic devices are often operated at room or higher temperature, calculations and measurements of temperature-dependent transport properties including magnetoresistance in a parallel magnetic field for different values of carrier density and QW width are very useful tool in determining the key scattering mechanisms and system parameters [6-14]. To the author's knowledge, up to now, no calculation of transport properties at finite temperature has been done for spin-polarized quasi-two-dimensional electron gas (Q2DEG) in a GaAs/InGaAs/GaAs QW. Therefore, the aim of this article is to calculate the finite temperature magnetoresistance of a Q2DEG realized in a GaAs/InGaAs/GaAs lattice-mismatched QW taking into account many-body effects which are very important when carrier density is very low [15-16]. We also calculate the zero-temperature total mobility and discuss the multiple-scattering effects which may lead to a metal–insulator transition (MIT) at low density [17-18].

## II. THEORY

We investigate a Q2DEG in the *xy* plane by using a realistic model of finitely deep quantum well. At low temperature, we assume that the electrons occupy only the lowest conduction subband. The exact envelope wave function for a finite square quantum well is given by [5]

$$\Psi(z) = C\sqrt{\frac{2}{L}} \begin{cases} \cos\left(\frac{1}{2}kL\right)\exp(\kappa z) & \text{for } z < 0 \\ \cos\left[k\left(z - \frac{1}{2}L\right)\right] & \text{for } 0 \leq z \leq L \\ \cos\left(\frac{1}{2}kL\right)\exp[-\kappa(z-L)] & \text{for } z > L \end{cases} \quad (1)$$

where $C$ is a normalization constant fixed by

$$C^2\left(1 + \frac{\sin a}{a} + \frac{1+\cos a}{b}\right) = 1 . \quad (2)$$

Here $a = kL$ and $b = \kappa L$ are dimensionless quantities given by the thickness of QW, $L$, and the barrier height $V_0$ as

$$a = \frac{L\sqrt{2m_z V_0}}{\hbar}\cos\left(\frac{1}{2}a\right) \quad (3)$$

and

$$b = a\tan\left(\frac{1}{2}a\right) \quad (4)$$

with $m_z$ being the effective mass of the electron along the growth direction.

When a parallel magnetic field $B$ is applied to the system, the Q2DEG is polarized and the electron densities $n_\pm$ at zero temperature for spin up/down are given as follow [6-8]

$$n_\pm = \frac{n}{2}\left(1 \pm \frac{B}{B_S}\right), \quad B < B_S, \quad (5)$$

$$n_+ = n, n_- = 0, \quad B \geq B_S.$$

Here, $n = n_+ + n_-$ is the total electron density and $B_S$ is the saturation field determined by $g\mu_B B_S = 2E_F$, where $\mu_B$ is the Bohr magneton, $g$ is the spin $g$-factor of electrons and $E_F$ is the Fermi energy, $E_F = \hbar^2 k_F^2 / 2m^*$, with $k_F = \sqrt{2\pi n}$ being the Fermi wave number and $m^*$ being the electron effective mass in *xy*-plane.

At finite temperature, $n_\pm$ has the form [8]

$$n_+ = \frac{n}{2}t\ln\frac{1 - e^{2x/t} + \sqrt{(e^{2x/t} - 1)^2 + 4e^{(2+2x)/t}}}{2}, \quad (6)$$

$$n_- = n - n_+$$

where $t = T/T_F$ with $T_F$ being the Fermi temperature and $x = B/B_S$. Note that the dependence of the carrier density on the magnetic field at arbitrary temperature can be obtained by minimizing the free energy (including Zeeman energy caused by the interaction of the magnetic field with the spin magnetic moment of the electrons) [19] with respect to spin polarization $(n_+-n_-)/n$. Within the approximation of noninteracting systems the analytical results can be obtained as shown in Eqs. 5 and 6. In the Boltzmann theory, the averaged transport relaxation time for the $(\pm)$ components is given as [7-8]

$$\langle \tau_{\pm}\rangle = \frac{\int d\varepsilon \tau(\varepsilon)\varepsilon\left[-\frac{\partial f^{\pm}(\varepsilon)}{\partial \varepsilon}\right]}{\int d\varepsilon \varepsilon\left[-\frac{\partial f^{\pm}(\varepsilon)}{\partial \varepsilon}\right]} \tag{7}$$

where $\varepsilon = \hbar^2 k^2/2m^*$ and

$$\frac{1}{\tau(\varepsilon)} = \frac{1}{(2\pi)^2 \hbar \varepsilon}\int_0^{2\pi}d\theta \int_0^{2k}\frac{\langle |U(\vec{q})|^2\rangle}{[\in(q,T)]^2}\frac{q^2 dq}{\sqrt{4k^2 - q^2}}, \tag{8}$$

$$\in(q,T) = 1 + \frac{2\pi e^2}{\varepsilon_L}\frac{1}{q}F_C(qL)[1-G(q)]\Pi(q,T), \tag{9}$$

$$\Pi(q,T) = \Pi_+(q,T) + \Pi_-(q,T), \tag{10a}$$

$$\Pi_{\pm}(q,T) = \frac{\beta}{4}\int_0^{\infty}d\mu'\frac{\Pi_{\pm}^0(q,\mu')}{\cosh^2[\frac{\beta}{2}(\mu_{\pm}-\mu')]}, \tag{10b}$$

$$\Pi_{\pm}^0(q,E_{F\pm}) = \Pi_{\pm}^0(q) = \frac{g_v m^*}{2\pi\hbar^2}\left[1 - \sqrt{1-\left(\frac{2k_{F\pm}}{q}\right)^2}\Theta(q-2k_{F\pm})\right], \tag{10c}$$

$$F_C(qL) = \frac{1}{C^4}\int_{-\infty}^{+\infty}dz\int_{-\infty}^{+\infty}dz'|\psi(z)|^2|\psi(z')|^2 e^{-q|z-z'|} \tag{11}$$

with $f^{\pm}(\varepsilon) = 1/\{1+\exp(\beta[\varepsilon-\mu_{\pm}(T)])\}$, $\beta = (k_B T)^{-1}$, $\mu_{\pm} = \ln[-1+\exp(\beta E_{F\pm})]/\beta$, $E_{F\pm} = \hbar^2 k_{F\pm}^2/(2m^*)$, $\vec{q} = (q,\theta)$, $\Theta(q)$ is the step function and $G(q)$ is a local-field correction (LFC) describing the many-body effects [15-16]. In the Hubbard approximation, only exchange effects are included and the LFC has the form $G_H(q) = q/[g_v g_s \sqrt{q^2+k_F^2}]$ where $g_v$ ($g_s$) is the valley (spin) degeneracy. To take into account both exchange and correlation effects, we also use analytical expressions $G_{GA}(q) = r_s^{4/3}1.402q/\sqrt{2.644C_{21}^2 q_s^2 + C_{22}^2 r_s^{4/3}q^2 - C_{23}r_s^{2/3}q_s q}$ where $r_s = 1/\sqrt{\pi a^{*2}n}$, $C_{2i}(r_s)$ ($i = 1, 2, 3$) are given in [16] and $q_s = g_s g_v / a^*$ with $a^* = \hbar^2 \varepsilon_L/(m^* e^2)$ as the effective Bohr radius. Here, $\varepsilon_L$ is the averaged dielectric constant of the system and $\langle |U(\vec{q})|^2\rangle$ is the random potential which depends on the scattering mechanism [15].

For the remote charged impurity scattering (RI), the random potential is given by [15]

$$\langle |U_{RI}(q)|^2\rangle = N_{RI}\left(\frac{2\pi e^2}{\varepsilon_L}\frac{1}{q}\right)^2 F_{RI}(q,z_i)^2 \tag{12}$$

where $N_{RI}$ is the 2D impurity density, $z_i$ is the distance of the impurities from the QW edge at $z = 0$, and $F_{RI}(q,z_i) = \int_{-\infty}^{+\infty}|\Psi(z)|^2 e^{-q|z-z_i|}dz$ is the form factor describing the electron-impurity interaction.

For the homogenous background impurity scattering (BI), the random potential has the form

$$\left\langle |U_{BI}(q)|^2 \right\rangle = \left(\frac{2\pi e^2}{\varepsilon_L}\frac{1}{q}\right)^2 \int_{-\infty}^{+\infty} dz_i N_i(z_i)[F_{RI}(q,z_i)]^2 = \left(\frac{2\pi e^2}{\varepsilon_L}\frac{1}{q}\right)^2 F_{BI}(q) \tag{13}$$

where $F_{BI}(q) = N_{B1}F_{B1}(q) + N_{B2}F_{B2}(q) + N_{B3}F_{B3}(q)$ with $F_{B1}(q) = \int_{-\infty}^{0}[F_{RI}(q,z_i)]^2 dz_i$, $F_{B2}(q) = \int_{0}^{L}[F_{RI}(q,z_i)]^2 dz_i$

and $F_{B3}(q) = \int_{L}^{\infty}[F_{RI}(q,z_i)]^2 dz_i$.

For the surface roughness scattering (SR), the random potential is given by [5]

$$\left\langle |U_{SR}(q)|^2 \right\rangle = \left(\frac{\pi^{1/2}\hbar^2 C^2 a^2 \Delta\Lambda}{m_z L^3}\right)^2 \exp(-q^2 \Lambda^2/4) \tag{14}$$

where $\Delta$ is the roughness amplitude and $\Lambda$ is the correlation length.

For the roughness-induced piezoelectric scattering (PESR), the random potential has the form [5]

$$\left\langle |U_{PE}(\bar{q})|^2 \right\rangle = \left(\frac{3\pi^{3/2} e e_{14} G A \varepsilon_{//} C^2 \Delta\Lambda}{8\varepsilon_L c_{44}}\right)^2 F_{PE}^2(t) \exp(-q^2 \Lambda^2/4)\sin^2 2\theta. \tag{15}$$

Here $\varepsilon_{//}$, $A$, $e_{ij}$, $c_{ij}$ are the lattice mismatch, anisotropy ratio, piezoelectric and elastic stiffness constants, respectively; $G = 2(2\frac{c_{12}}{c_{11}}+1)(c_{11}-c_{12})$ and $F_{PE}(q,z;L)$ is the form factor for the piezoelectric field [5],

$$F_{PE}(q,z;L) = \frac{1}{2q}\begin{cases} e^{qz}(1-e^{-2qL}) & , z < 0 \\ e^{-qz}(1+2qz) - e^{-q(2L-z)} & , 0 \leq z \leq L \\ 2qLe^{-qz} & , z > L \end{cases} \tag{16}$$

The mobility of the unpolarized and fully polarized 2DEG can be calculated as $\mu = e\langle\tau\rangle/m^*$. The resistivity can be obtained using the relation $\rho = 1/\sigma$ where $\sigma = \sigma_+ + \sigma_-$ is the total conductivity with $\sigma_\pm$ as the conductivity of the $(\pm)$ spin component given by $\sigma_\pm = n_\pm e^2 \langle\tau_\pm\rangle/m^*$ [7-8].

To determine the total mobility limited by the SR, PESR, RI and BI scattering we can use the Matthiessen's rule,

$$\frac{1}{\langle\tau_{tot}\rangle} = \frac{1}{\langle\tau_{SR}\rangle} + \frac{1}{\langle\tau_{PESR}\rangle} + \frac{1}{\langle\tau_{RI}\rangle} + \frac{1}{\langle\tau_{BI}\rangle}. \tag{17}$$

It is well-known that at low electron densities interaction effects become inefficient to screen the random potential and the MIT can be occurred. The MIT can be explained by taking into account the multiple-scattering effect (MSE). The MIT is then described by parameter $A_o$ [14, 17-18],

$$A_o = \frac{1}{8\pi^2 n^2} \int_0^\infty \int_0^{2\pi} \frac{\langle |U(\bar{q})^2|\rangle [\Pi^0(q)]^2 q\, dq\, d\theta}{[\varepsilon(q)]^2}. \tag{18}$$

For $n > n_{MIT}$, where $A_o < 1$, the Q2DEG is in a metallic phase and the mobility $\mu_{MSE}$ can be obtained using the following approximated relation [20]

$$\mu_{MSE} = \mu(1 - A_o). \qquad (19)$$

For $n < n_{MIT}$, where $A_o > 1$, the Q2DEG is in an insulating phase and $\mu_{MSE} = 0$.

## III. NUMERICAL RESULTS

We have performed numerical calculations of the resistance ratio $\rho(B_s)/\rho(B = 0)$ and the zero-temperature total mobility, taking into account the many-body and multiple-scattering effects. We use $V_0 = 131$ meV and $m_z = m^* = 0.058 m_0$, where $m_0$ is the free electron mass.

### III.1. The resistance ratio $\rho(B_s)/\rho(B = 0)$ for SR and PESR scattering

The resistance ratio $\rho(B_s)/\rho(B = 0)$ as a function of electron density is shown in Fig. 1 for $L = 100$Å, $\Delta = 5$ Å, $\Lambda = 50$ Å and different LFC models. For $T = 0$ we observe that the resistance ratio decreases with the increase in electron density. At low (high) densities, we find that the resistivity of a polarized 2DEG is higher (lower) in comparison with that of the unpolarized case and the LFC effects are considerable (negligible). For $T \sim 0.3 T_F$ the resistance ratio is lower (higher) than that of zero-temperature case at low (high) densities.

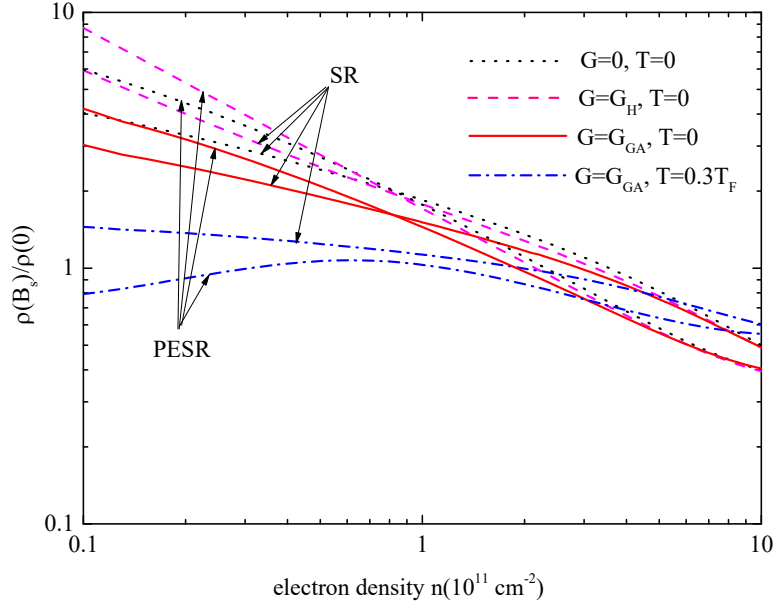

**Fig. 1.** Resistance ratio $\rho(B_s)/\rho(B = 0)$ as a function of electron density for SR and PESR scattering for $\Delta = 5$ Å, $\Lambda = 50$ Å, $L = 100$ Å and three $G(q)$ models.

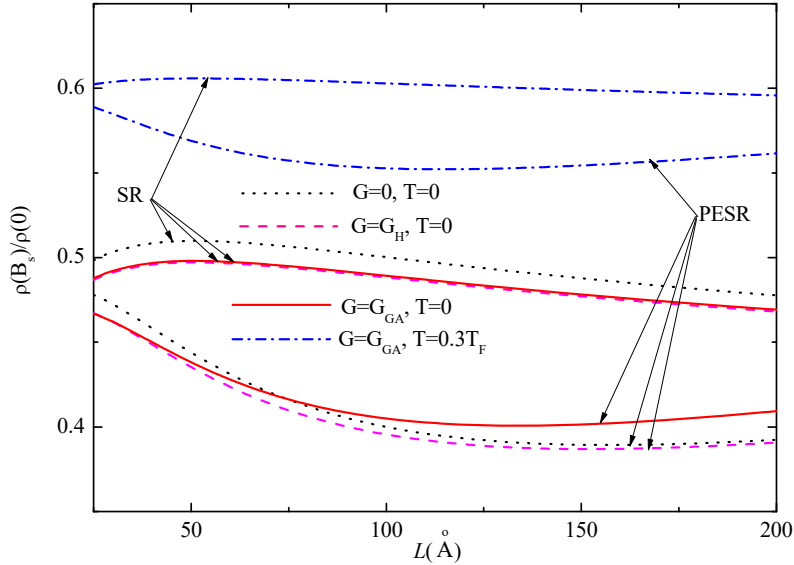

**Fig. 2.** Resistance ratio $\rho(B_s)/\rho(B = 0)$ versus QW width $L$ for SR and PESR at temperature $T = 0$ and $T = 0.3T_F$ for $n = 10^{12}$ cm$^{-2}$, $\Delta = 5$ Å, $\Lambda = 50$ Å and three $G(q)$ models.

The resistance ratio as a function of QW width $L$ for SR and PESR scattering at temperature $T = 0$ and $T = 0.3T_F$ is plotted in Fig. 2 for $n = 10^{12}$ cm$^{-2}$, $\Delta = 5$ Å, $\Lambda = 50$ Å and different $G(q)$ models. We see that, for entire range of QW width considered, the resistivity of a unpolarized 2DEG is higher in comparison with the polarized case, the many-body effect is negligible for SR scattering, and the temperature effect is always notable. For $L > 125$ Å the dependence of resistance ratio on QW width is very weak.

### III.2. The resistance ratio $\rho(B_s)/\rho(B = 0)$ for BI and RI scattering

Resistance ratio $\rho(B_s)/\rho(B = 0)$ versus electron density for BI and RI scattering at $T = 0$ and $T = 0.3T_F$ is plotted in Fig. 3 for $L = 100$ Å, $N_{B1} = N_{B2} = N_{B3} = 10^{17}$ cm$^{-3}$ and $N_{RI} = n$ in different approximations for the LFC. The distance between 2DEG and remote impurities $z_i$ is assumed to be $-L/2$. It is seen that the resistance ratio decreases with the increase in electron density. This behavior has been explained by Dolgopolov and Gold, using a qualitative calculation of the scattering time [10]. For $T = 0$ we observe that at low densities the resitivity of a polarized 2DEG is higher in comparison with that of the unpolarized case due to an enhancement of the 2D Fermi wave vector and a suppression of the effective 2D screening wave vector in the parallel magnetic field. The LFC affects remarkably the resistance ratio because the exchange-correlation effects are very important at low densities. For $T = 0.3T_F$ the resistance ratio for BI scattering is lower (higher) than that for $T = 0$ at low (high) densities. The finite temperature resistivity of a polarized 2DEG is always lower in comparison with that of the unpolarized case for entire density range considered for both BI and RI scattering.

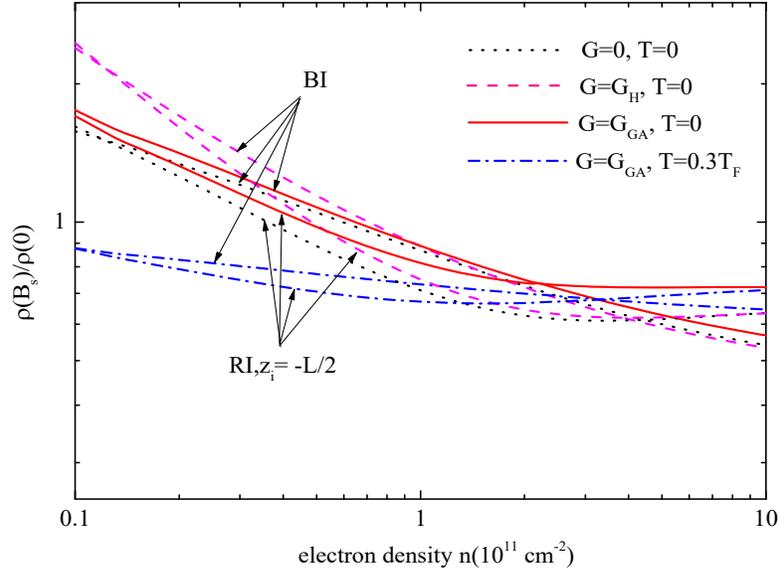

**Fig. 3.** Resistance ratio $\rho(B_s)/\rho(B = 0)$ versus electron density for BI and RI scattering at $T = 0$ and $T = 0.3T_F$ for $L = 100$ Å, $N_{B1} = N_{B2} = N_{B3} = 10^{17}$ cm$^{-3}$ and $N_{RI} = n$ in different approximations for the LFC.

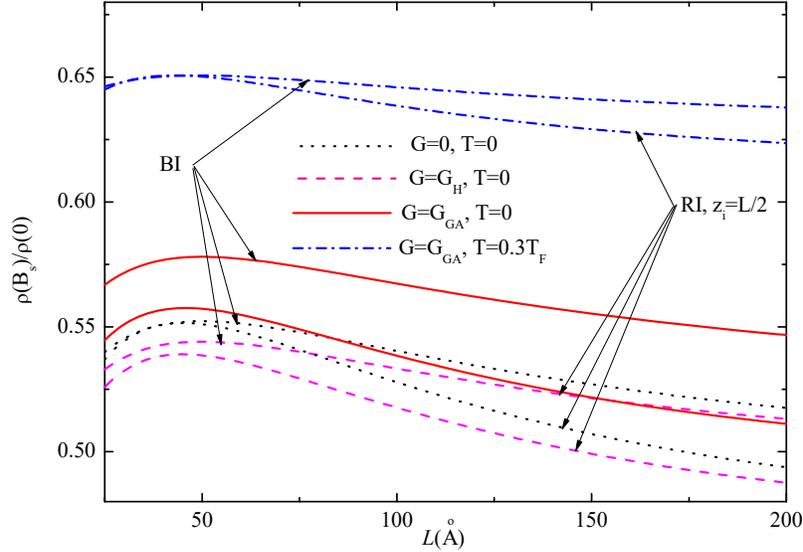

**Fig. 4.** Resistance ratio $\rho(B_s)/\rho(B = 0)$ versus QW width $L$ for BI and RI scattering at temperature $T = 0$ and $0.3T_F$ for $N_{B1} = N_{B2} = N_{B3} = 10^{17}$ cm$^{-3}$ and $N_{RI} = n = 10^{12}$ cm$^{-2}$ in three $G(q)$ models.

In Fig.4, we plot the resistance ratio $\rho(B_s)/\rho(B = 0)$ versus QW width $L$ for BI and RI scattering at temperature $T = 0$ and $0.3T_F$ for $N_{B1} = N_{B2} = N_{B3} = 10^{17}$ cm$^{-3}$ and $N_{RI} = n = 10^{12}$ cm$^{-2}$ in three $G(q)$ models. The remote impurities are assumed to be in the middle of the QW. We see that the resistance ratio increases with QW width $L$, reaches a peak and then decreases with further increase in $L$. The resistivity of a unpolarized 2DEG is higher in comparison with that of a polarized one for all $L$ values considered. The differences between the results of LFC models are considerable and the resistance ratio $\rho(B_s)/\rho(B = 0)$ increases substantially with temperature.

### III.3. The total mobility and multiple-scattering effects

The zero-field and zero-temperature total mobility, limited by SR, PESR, BI and RI scattering, versus electron density $n$ for $L = 100$ Å, $\Delta = 5$ Å, $\Lambda = 50$ Å, $N_{B1} = N_{B2} = N_{B3} = 10^{17}$ cm$^{-3}$, $N_{RI} = n$ and $z_i = -L/2$ in two $G(q)$ models is plotted in Fig 5. At low density, we see that the many-body and multiple-scattering effects are considerable. At high density ($n > 10^{12}$ cm$^{-2}$) the MSE is not important and the total mobility in $G_H$ and $G_{GA}$ model is almost identical. The critical density $n_{MIT}$ for MIT is about 2. $10^{11}$ cm$^{-2}$ and its value in case of $G_{GA}$ model is somewhat smaller than that obtained by using Hubbard approximation. Finally, we note that, at low densities, the simple approximation for $\mu_{MSE}$ given in Eq. 19 gives results very close to those obtained by more complicated self-consistent multiple-scattering theory [21-23].

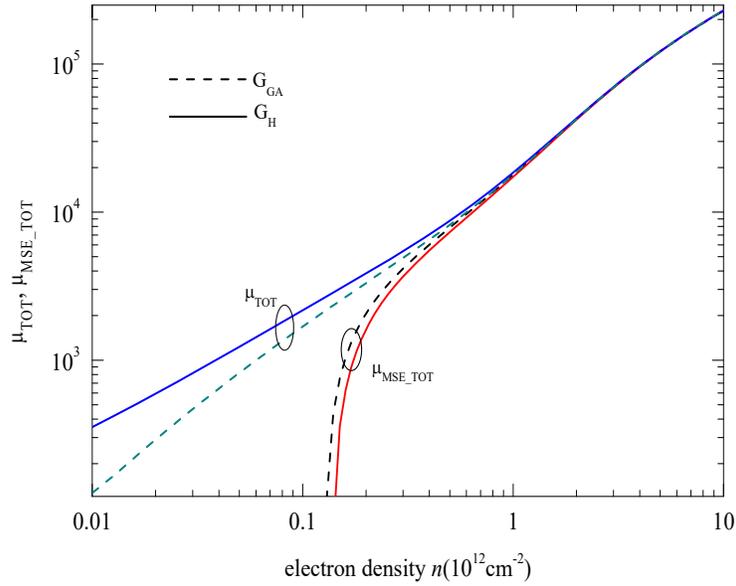

**Fig. 5**. Total mobility, limited by SR, PESR, BI and RI scattering, versus electron density for $L = 100$ Å $\Delta = 5$ Å, $\Lambda = 50$ Å, $N_{B1} = N_{B2} = N_{B3} = 10^{17}$ cm$^{-3}$, $N_{RI} = n$, and $z_i = -L/2$ in two $G(q)$ models.

### IV. CONCLUSION

In conclusion, we have performed the calculation of the resistance ratio $\rho(B_s)/\rho(B = 0)$ as a function of electron density $n$ and QW width $L$ in three $G(q)$ models at zero and finite temperatures for four scattering mechanisms: SR, PESR, RI and BI. We find the remarkable difference between the results of $G = 0$, $G_H$, and $G_{GA}$ models at low densities. For all scattering mechanisms considered, the temperature and magnetic field effects are remarkable for the entire range of QW width, especially at low densities. For wide QWs the dependence of resistance ratio on QW width is relatively weak. We have also calculated the zero-field and zero-temperature total mobility as a function of carrier density $n$ and shown that the MSE leads to a MIT at low density. We find that the critical density $n_{MIT}$ for MIT in case of $G_{GA}$ model is somewhat smaller than that obtained by using Hubbard approximation. The dependence of the resistivity on magnetic field, $n$, $L$, temperature and LFC shown in this paper can be used in combination with possible future measurements to get information about the scattering mechanisms and many-body effects in GaAs/InGaAs/GaAs lattice-mismatched QW structures [14].

**ACKNOWLEDGEMENT**

This research is funded by Vietnam National Foundation for Science and Technology Development (NAFOSTED) under Grant number 103.01-2017.23.